\documentclass[12pt]{article}
\usepackage{a4wide}
\usepackage{amssymb}
\usepackage{graphicx}
\usepackage{caption}
\usepackage{subcaption}
\usepackage{mathdots}
\usepackage{color}
\usepackage[all]{xy}
\usepackage{bbm}
\usepackage[title, titletoc]{appendix}
\usepackage{amsmath}

\usepackage{comment}

\def\drawbox#1#2{\hrule height#2pt 
        \hbox{\vrule width#2pt height#1pt \kern#1pt 
              \vrule width#2pt}
              \hrule height#2pt}

\def\Fund#1#2{\vcenter{\vbox{\drawbox{#1}{#2}}}}
\def\Asym#1#2{\vcenter{\vbox{\drawbox{#1}{#2}
              \kern-#2pt       
              \drawbox{#1}{#2}}}}

\def\funda{\Fund{6.5}{0.4}}

\def\symm{\funda\kern-0.4pt\funda}

\def\makeatletter{\catcode`\@=11}
\makeatletter
\def\mathbox#1{\hbox{$\m@th#1$}}%
\def\math@ccstyles#1#2#3#4#5#6#7{{\leavevmode
      \setbox0\mathbox{#6#7}%
      \setbox2\mathbox{#4#5}%
      \dimen@ #3%
      \baselineskip\z@\lineskiplimit#1\lineskip\z@
      \vbox{\ialign{##\crcr
             \hfil \kern #2\box2 \hfil\crcr
             \noalign{\kern\dimen@}%
             \hfil\box0\hfil\crcr}}}}
\def\mathaccstyles{\math@ccstyles\maxdimen}
\def\maththroughstyles{\math@ccstyles{-\maxdimen}}
\def\unity%
 {\maththroughstyles{.45\ht0}\z@\displaystyle {\mathchar"006C}\displaystyle 1}

\def\be{\begin{eqnarray}}
\def\ee{\end{eqnarray}}

\begin{document}

\setcounter{table}{0}

\mbox{}
\vspace{2truecm}
\linespread{1.1}

\begin{center}
{\LARGE \bf 2d duality for orthogonal gauge theories with 8 supersymmetries}

\end{center}

\vspace{2truecm}

\centerline{
    {    {\large \bf Eran Avraham${}^{a}$} \footnote{eranav@post.bgu.ac.il}
   {\bf and}    
    \large \bf Oren Bergman ${}^{a}$} \footnote{bergman@physics.technion.ac.il}
}

 \vspace{1cm}
 \centerline{{\it ${}^a$ Department of Physics, Technion, Israel Institute of Technology}} \centerline{{\it Haifa, 32000, Israel}}
 \vspace{1cm}

 \centerline{\bf ABSTRACT}
  \vspace{1truecm}
We recently conjectured a set of dualities relating two-dimensional orthogonal gauge theories with ${\cal N}=(4,4)$ supersymmetry,
analogous to Hori's dualities with ${\cal N}=(2,2)$ supersymmetry.
Here we provide a quantitative test of this conjecture by computing the elliptic genera of the dual pairs and showing that they agree.
The elliptic genus of orthogonal gauge theories has multiple topological sectors that depend on the global structure of the group
and on the value of a discrete $\theta$ parameter.
We derive the dependence on the $\theta$ parameter by determining whether a given sector has $(S)Pin$ structure or not.


 \vspace{1truecm}

 \newpage

\tableofcontents

\section{Introduction}

Two dimensional supersymmetric gauge theories have rich dynamics and exhibit IR dualities similar to their cousins in higher dimensions.
In fact, there is a web of connections between the various dualities that goes all the way to six dimensions upon compactification.
Deciphering this underlying structure is a long standing goal.
A common theme is to use localization techniques to verify and solidify geometric intuition coming from string theory and M-theory realizations. 
Based on a such a string theory realization, we have recently conjectured a set of dualities relating two dimensional
${\cal N}=(4,4)$ supersymmetric gauge theories with orthogonal and symplectic gauge groups \cite{Bergman:2018vqe}, 
generalizing a conjecture relating unitary gauge theories made by Brodie in \cite{Brodie:1997wn}.
In this work, we implement the elliptic genus machinery of \cite{Benini:2013nda,Benini:2013xpa} to further study these dualities.

Gauge theories in two dimensions based on the $so(k)$ algebra are particularly interesting and subtle for two reasons. 
First there are several possibilities for the
global structure of the group, including $SO(k)$, $O(k)_\pm$, $Spin(k)$ and $Pin^{\pm}(k)$, which give rise to distinct gauge theories,
with a distinct spectrum of local and line operators
\cite{Aharony:2013hda,Cordova:2017vab}.\footnote{The $O(k)_\pm$ theories correspond to $\mathbb{Z}_2$ orbifolds of the $SO(k)$
theory by charge conjugation and by charge conjugation combined with $(-1)^F$, respectively \cite{Hori:2011pd}.}
Second, except for $Spin(k)$, the theories may possess multiple topological sectors that contribute in the path integral.
For example an $SO(k)$ theory (with $k>2$) on $S^2$ has two topological sectors corresponding to monopole number $n\in\{0,1\}$,
weighted by a phase given by $\exp{in\theta}$, where $\theta$ is a discrete analog of a theta parameter taking values in $\{0,\pi\}$.
Investigating these theories therefore requires one to classify the different gauge bundles on the given spacetime manifold. 

Orthogonal gauge heories with ${\cal N}=(2,2)$ supersymmetry and fundamental chiral multiplets have been argued to satisfy a set of Seiberg-like 
dualities given by \cite{Hori:2011pd}
\be
SO(k) + Q_i &\longleftrightarrow & O(N-k+1)_+ + q_i + S_{ij} \\
O(k)_+ + Q_i &\longleftrightarrow & SO(N-k+1) + q_i + S_{ij}\\
O(k)_-  + Q_i & \longleftrightarrow & O(N-k+1)_- + q_i + S_{ij}\,,
\ee
where $i=1,\ldots ,N$, and the theory on the RHS also has a gauge singlet $S$ in the symmetric representation of the flavor $U(N)$ symmetry,
and a superpotential $W = qSq$.
The value of the discrete theta parameter $\theta$ is fixed in each case such that the Coulomb branch is lifted.
These dualities were further tested by \cite{Closset:2017vvl,Kim:2017zis}.
In particular the $T^2$ partition functions, or elliptic genus, of the dual theories for $k\leq 4$ were computed, and found to agree, in \cite{Kim:2017zis}.
A brane realization of these dualities in M theory was subsequently provided in \cite{Bergman:2018vqe}.

Interestingly, a very similar brane construction suggested an analogous set of dualities for ${\cal N}=(4,4)$ theories 
with fundamental hypermultiplets given roughly by
\be
(S)O(k)  + Q_i & \longleftrightarrow & (S)O(2n-k+1) + q_i \,,
\ee
where $i=1,\ldots,n$. There are no extra singlets in this case, and $\theta = \pi$ on both sides.
The brane construction does not seem to differentiate $O(k)$ from $SO(k)$, or $O(k)_+$ from $O(k)_-$,
so we were not able to make a more refined set of conjectures as in the ${\cal N}=(2,2)$ case.
More concretely, the ${\cal N}=(4,4)$ duality is the statement that the two gauge theories flow to the same superconformal theory on the Higgs branch.
This is supported in part by the fact that the central charges of the ``magnetic" and ``electric" Higgs branch SCFT's agree:
\be
\widehat{c}_m = (2n-k+1)n - \frac{1}{2}(2n-k+1)(2n-k)  = kn - \frac{1}{2}k(k-1) = \widehat{c}_e \,.
\ee
The aim of this work is to provide further evidence for, and to make more precise, the ${\cal N}=(4,4)$ dualities of orthogonal theories
via the elliptic genus.

In section \ref{flat conn} we recall the structure of flat $O(k)$ bundles on the torus, and classify them according to whether or not they
admit $Pin^+$ structure, which is relevant for the dependence on the discrete theta parameter.
Then in section \ref{(4,4) theories} we test the ${\cal N}=(4,4)$ dualities for orthogonal gauge theories by comparing the elliptic genera
for some low rank cases. We also test a twisted generalization of the dualities and show that the three pairs of dualities are connected by gauging of a global $\mathbb{Z}_2$ symmetry. 
In the appendix we briefly review the elliptic genus formulas and then analytically compare the elliptic genera of the lower rank dualities using the known identities of the Jacobi-theta functions.

\section{Flat $O(k)$ connections on $T^2$}
\label{flat conn}
Flat $O(k)$ connections on $T^2$ were classified in \cite{Kim:2014dza}.
There are eight sectors in general given by the following pairs of Wilson lines along the temporal and spatial circles of the torus, $(U_1,U_2)$. 
For $O(2p)$ with $p\geq 2$:
\be
\label{O(2p)Connections+}
(U_1,U_2)^+_{ab} &=& \left\{
\begin{array}{l}
({\rm diag}(e^{iu_{1i}\sigma_2}),{\rm diag}(e^{iu_{2i}\sigma_2}))\\
({\rm diag}(e^{iu_{1i}\sigma_2},1,1),{\rm diag}(e^{iu_{2i}\sigma_2},1,-1))\\
({\rm diag}(e^{iu_{1i}\sigma_2},1,-1),{\rm diag}(e^{iu_{2i}\sigma_2},1,1))\\
({\rm diag}(e^{iu_{1i}\sigma_2},1,-1),{\rm diag}(e^{iu_{2i}\sigma_2},1,-1))
\end{array}
\right. \\[10pt]
\label{O(2p)Connections-}
(U_1,U_2)^-_{ab} &=& \left\{
\begin{array}{l}
({\rm diag}(e^{iu_{1i}\sigma_2},1,-1,-1,1),{\rm diag}(e^{iu_{2i}\sigma_2},1,1,-1,-1))\\
({\rm diag}(e^{iu_{1i}\sigma_2},-1,-1),{\rm diag}(e^{iu_{2i}\sigma_2},1,-1))\\
({\rm diag}(e^{iu_{1i}\sigma_2},1,-1),{\rm diag}(e^{iu_{2i}\sigma_2},-1,-1))\\
({\rm diag}(e^{iu_{1i}\sigma_2},1,-1),{\rm diag}(e^{iu_{2i}\sigma_2},-1,1))\,,
\end{array}
\right.
\ee
where $(a,b)=(0,0),(0,1),(1,0)$ and $(1,1)$, respectively,
and for $O(2p+1)$ with $p\geq 1$:
\be
\label{O(2p+1)Connections+}
(U_1,U_2)^+_{ab} &=& \left\{
\begin{array}{l}
({\rm diag}(e^{iu_{1i}\sigma_2},1),{\rm diag}(e^{iu_{2i}\sigma_2},1))\\
({\rm diag}(e^{iu_{1i}\sigma_2},1),{\rm diag}(e^{iu_{2i}\sigma_2},-1))\\
({\rm diag}(e^{iu_{1i}\sigma_2},-1),{\rm diag}(e^{iu_{2i}\sigma_2},1))\\
({\rm diag}(e^{iu_{1i}\sigma_2},-1),{\rm diag}(e^{iu_{2i}\sigma_2},-1))
\end{array}
\right. \\[10pt]
\label{O(2p+1)Connections-}
(U_1,U_2)^-_{ab} &=& \left\{
\begin{array}{l}
({\rm diag}(e^{iu_{1i}\sigma_2},-1,-1,1),{\rm diag}(e^{iu_{2i}\sigma_2},1,-1,-1))\\
({\rm diag}(e^{iu_{1i}\sigma_2},-1,-1,1),{\rm diag}(e^{iu_{2i}\sigma_2},1,-1,1))\\
({\rm diag}(e^{iu_{1i}\sigma_2},1,-1,1),{\rm diag}(e^{iu_{2i}\sigma_2},-1,-1,1))\\
({\rm diag}(e^{iu_{1i}\sigma_2},1,1,-1),{\rm diag}(e^{iu_{2i}\sigma_2},1,-1,1))\,.
\end{array}
\right.
\ee
For $O(1)$ there are only the four sectors in (\ref{O(2p+1)Connections+}),
\be
  (U_1,U_2)^{+}_{ab}=(1,1),\ (1,-1),\ (-1,1),\ (-1,-1)\,,
\ee
and for $O(2)$ the first sector in (\ref{O(2p)Connections-}) is absent, leaving only seven sectors,
\be
\label{O(2)Connections}
  (U_1,U_2)^{+}_{ab}&=&(e^{iu_1\sigma_2},e^{iu_2\sigma_2}), \, (\mathbbm{1},\sigma_3), \, (\sigma_3,\mathbbm{1}), \,(\sigma_3,\sigma_3) \\
 (U_1,U_2 )^-_{01,10,11} &=&(-\mathbbm{1},\sigma_3),(\sigma_3,-\mathbbm{1}), (\sigma_3,-\sigma_3)\,. \nonumber
\ee

\subsection{$Pin$ structure}
\label{Pin structure}

An $O(k)$ bundle over $X$ lifts to a $Pin^\pm(k)$ bundle over $X$ provided that $w_2=0$ and $w_2 + w_1^2=0$, respectively,
where $w_1$ and $w_2$ are the first and second Stiefel-Whitney classes of the $O(k)$ bundle.
We are specifically interested in bundles over $T^2$ with a flat connection.
The corresponding sector in the path integral is accompanied by a phase $e^{i\theta w_2}$, where $\theta\in \{0,\pi\}$ is the discrete theta parameter.
We therefore concentrate on the case of $Pin^+(k)$.
Namely we would like to classify flat $O(k)$ connections on $T^2$ according to whether they lift to flat $Pin^+(k)$ connections on $T^2$.

A closely related problem was considered in the context of the toroidal compactification of Type I string theory by Witten in \cite{Witten:1997bs}.
Since the precise gauge symmetry of Type I string theory is $Spin(32)/\mathbb{Z}_2$, one can consider a compactification on $T^2$ 
with a flat connection that does not lift to a flat $Spin(32)$ connection. This was called ``toroidal compactification without vector structure".
The condition for the existence of vector structure is $\tilde{w}_2=0$, where $\tilde{w}_2$ is defined in an analogous manner to $w_2$.
A flat $Spin(32)/\mathbb{Z}_2$ connection on $T^2$ is given by a pair $U_1,U_2$ satisfying $U_1 U_2 = U_2 U_1$.
The lift to $Spin(32)$ then satisfies $U_1 U_2 = (-1)^{\tilde{w}_2} U_2 U_1$.
For our case, flat $O(k)$ connections on $T^2$ lift to $Pin^+(k)$ connections satisfying $U_1 U_2 = (-1)^{w_2} U_2 U_1$.

We will show that the flat $O(k)$ connections $(U_1,U_2)^{+}$ correspond to bundles with $Pin^+$ structure, namely $w_2=0$,
whereas $(U_1,U_2)^{-}$ correspond to bundles without $Pin^+$ structure, namely $w_2\neq 0$.
The latter will therefore contribute to the $T^2$ partition function with a phase $e^{i\theta}$.


Let us recall the definition of $Pin(k)$ \cite{Atiyah:1964zz}.
The group $Pin(k)$ is a double cover of the group $O(k)$, in the same way that $Spin(k)$ is a double cover of $SO(k)$.
There is a two-to-one map $\rho: Pin(k) \rightarrow O(k)$ that takes an element $\Lambda\in Pin(k)$ into the element $R\in O(k)$ given by
\be
R_{ij}=\pm \frac{1}{k}\text{Tr}(\Lambda \gamma_i \Lambda^\dagger \gamma_j) \,,
\ee
where the sign depends on whether $\Lambda$ consists of an even or odd number of gamma matrices.
More explicitly $Pin^{\pm}(k)$ is generated by $S_{ij}=\frac12[\gamma_i,\gamma_j]$, together with a parity operator $P$
that satisfies $\mbox{det}(\rho(P)) = -1$, and $P^2=\pm 1$.
For $Pin^+(k)$ we can take $P=\gamma_i$ for some $i$ for even $k$, and $P=\prod_{i=1}^k \gamma_i$ for odd $k$.
In these cases $\rho(P) = \mbox{diag}(+1,\ldots,-1,\ldots,+1)$ for even $k$ and 
$\rho(P)=\mbox{diag}(-1,\ldots,-1)$ for odd $k$.

Let us first consider the $k=2,3,4$ cases explicitly, and then generalize to all $k$.
For $k=2$ we can choose $\gamma_1=\sigma_1$, $\gamma_2=\sigma_2$, and $P=\gamma_1 = \sigma_1$.
The map $Pin^+(2) \rightarrow O(2)$ takes the form
\be
\Lambda(\phi)=e^{i\frac{\phi}{2}S}=
\left(
\begin{array}{ccccccccc}
e^{i\frac{\phi}{2}} & 0  \\
0 & e^{-i\frac{\phi}{2}} 
\end{array}
\right)&\mapsto & \rho(\Lambda) = R(\phi)=
\left(
\begin{array}{ccccccccc}
cos(\phi) & sin(\phi)  \\
-sin(\phi) & cos(\phi) 
\end{array}
\right)\\
P=\sigma_1&\mapsto & \rho(P)=\left(
\begin{array}{ccccccccc}
-1 & 0  \\
0 & 1
\end{array}
\right) \,.
\ee
Using these relations one can easily find lifts to $Pin^+(2)$ of the commuting flat $O(2)$ connections in (\ref{O(2)Connections}).
For $(U_1,U_2)^{+}$ we have
\be
(\Lambda(u_1),\Lambda(u_2)) &\mapsto & (e^{iu_1\sigma_2},e^{iu_2\sigma_2}) \nonumber \\
(\mathbbm{1},P\Lambda(\pi)) &\mapsto & (\mathbbm{1},\sigma_3) \nonumber \\
(P\Lambda(\pi),\mathbbm{1})) &\mapsto & (\sigma_3,\mathbbm{1})  \\
(P\Lambda(\pi),P\Lambda(\pi)) &\mapsto & (\sigma_3,\sigma_3) \nonumber \,.
\ee
The $Pin^+(2)$ holonomy pairs on the LHS all commute and so correspond to $w_2=0$.
For $(U_1,U_2)^{-}$ on the other hand we find
\be
(\Lambda(\pi),P\Lambda(\pi)) &\mapsto & (-\mathbbm{1},\sigma_3) \nonumber \\
(P\Lambda(\pi),\Lambda(\pi)) &\mapsto & (\sigma_3,-\mathbbm{1}) \nonumber \\
(P\Lambda(\pi),P) &\mapsto & (\sigma_3,-\sigma_3) \,,
\ee
which anti-commute in $Pin^+(2)$ and therefore correspond to $w_2\neq 0$.
For $k=3$ we can take $\gamma_1=\sigma_1\otimes\mathbbm{1}$, $\gamma_2=\sigma_3\otimes\mathbbm{1}$, 
$\gamma_3= \sigma_2\otimes \sigma_2$, and 
$P=\gamma_1 \gamma_2 \gamma_3 = \mathbbm{1}\otimes \sigma_2$.
In this case $\rho(P) = \mbox{diag}(-1,-1,-1)$.
The lifts to $Pin^+(3)$, up to signs, of the $(U_1,U_2)^{+}$ holonomies in (\ref{O(2p+1)Connections+}) are given by the commuting pairs
\be
(\Lambda_{12}(u_1),\, \Lambda_{12}(u_2)),\, (\Lambda_{12}(u_1),P\Lambda_{12}(u_2)),\,
(P\Lambda_{12}(u_1),\Lambda_{12}(u_2)),\, (P\Lambda_{12}(u_1),P\Lambda_{12}(u_2)) \,,
\ee
and the lifts of $(U_1,U_2)^{-}$ are given by the anti-commuting pairs
\be
(\Lambda_{12}(\pi),\Lambda_{23}(\pi)) ,\,
(\Lambda_{12}(\pi),P\Lambda_{31}(\pi)), \,
(P\Lambda_{31}(\pi),\Lambda_{12}(\pi)) ,\,
(P\Lambda_{12}(\pi),P\Lambda_{31}(\pi))\,.
\ee 
As before, the former correspond to $w_2=0$ and the latter to $w_2\neq 0$.
For $k=4$ we can take
$\gamma_1=i\sigma_2\otimes i\sigma_2\otimes\mathbbm{1}$, $\gamma_2=i\sigma_2\otimes\sigma_1\otimes i\sigma_2$, 
$\gamma_3=i\sigma_2\otimes\sigma_3\otimes i\sigma_2$,
$\gamma_4=\sigma_3 \otimes \mathbbm{1} \otimes \mathbbm{1}$,
and $P=\gamma_4$.
In this case $\rho(P) = \mbox{diag}(1,1,1,-1)$.
The $O(4)$ pairs $(U_1,U_2)^{+}$ lift to the commuting pairs
\be
(\Lambda_{12}(u_{11})\Lambda_{34}(u_{12}),\Lambda_{12}(u_{21})\Lambda_{34}(u_{22})), &
(\Lambda_{12}(u_{11}),P\Lambda_{12}(u_{21})), \\
(P\Lambda_{12}(u_{11}),\Lambda_{12}(u_{21})), &
(P\Lambda_{12}(u_{11}),P\Lambda_{12}(u_{21})) \,,
\ee
and the pairs $(U_1,U_2)^{-}$ lift to the anti-commuting pairs
\be
(\Lambda_{23}(\pi),\Lambda_{34}(\pi)) , &
(\Lambda_{12}(u_1)\Lambda_{34}(\pi),P\Lambda_{12}(u_2)) , \\
(P\Lambda_{12}(u_1),\Lambda_{12}(u_2)\Lambda_{34}(\pi)) , &
(P\Lambda_{12}(u_1),P\Lambda_{12}(u_2)\Lambda_{34}(\pi)) \,.
\ee

The generalization to higher $k$ is straightforward.
In (\ref{O(2p)Connections+})-(\ref{O(2p+1)Connections-}) we see that the $O(2p)$ and $O(2p+1)$ 
bundles decompose into a direct sum of commuting flat connections of 
$SO(2)^{p-1}$ and those of $O(4)$ and $O(3)$, respectively.
The first part lifts trivially to a flat $Spi n(2)^{p-1}$ bundle, since $Spin(2)=SO(2)$.
To conclude, the pairs $(U_1,U_2)^+$ define flat $O(k)$ bundles on $T^2$ with trivial $w_2$,
while the pairs $(U_1,U_2)^-$ define flat $O(k)$ bundles on $T^2$ with non-trivial $w_2$.

\section{${\cal N}=(4,4)$ theories}
\label{(4,4) theories}

Two dimensional gauge theories with ${\cal N}=(4,4)$ supersymmetry have a moduli space of vacua that generically has two branches,
a Coulomb branch and a Higgs branch.
The two branches decouple in the IR and the theory flows to two independent superconformal field theories with an $SU(2)\times SU(2)$
R-symmetry \cite{Witten:1997yu}.
In some cases the Coulomb branch is lifted and one is left with just the Higgs branch SCFT.
In particular this is the case for $O(k)$ or $SO(k)$ with any number $N$ of fundamental hypermultiplets and with 
$\theta = \pi$ \cite{Bergman:2018vqe}.
This is somewhat surprising given the fact that in theories with eight supersymmetries in higher dimensions the Coulomb branch cannot be lifted.
It is in fact very similar to what happens in ${\cal N}=(2,2)$ supersymmetric gauge theories \cite{Hori:2011pd},
and is due partly to an effective linear twisted superpotential on the Coulomb branch.\footnote{The twisted superpotential lifts the part
of the Coulomb branch corresponding to the scalar in the ${\cal N}=(2,2)$ vector multiplet. Presumably a linear superpotential is also
generated for the adjoint chiral multiplet, though this has not been demonstrated yet.}
We will also see that it is consistent with the results we will present below: the Witten index is a finite integer, 
as it should be for a regular theory. 

As stated in the introduction, the brane construction of \cite{Bergman:2018vqe} suggests a duality between the $O(k)$ (or $SO(k)$)
theory with $n$ fundamental hypermultiplets and the $O(2n-k+1)$ (or $SO(2n-k+1$) theory with $n$ fundamental hypermultiplets.
We will provide more evidence for this conjecture by comparing the elliptic genera.
Our analysis shows that, as in the ${\cal N}=(2,2)$ theories, there are actually three dualities between ${\cal N}=(4,4)$ theories:
\be
SO(k) + Q_i &\longleftrightarrow & O(2n-k+1)_+ + q_i\\ [5pt]
O(k)_+ + Q_i &\longleftrightarrow & SO(2n-k+1) + q_i \\[5pt]
O(k)_- + Q_i &\longleftrightarrow & O(2n-k+1)_- + q_i \,,
\ee
where $\theta = \pi$ in all cases.


To compute the elliptic genus of the ${\cal N}=(4,4)$ theories we use the results of \cite{Benini:2013nda,Benini:2013xpa} for the 
${\cal N}=(2,2)$ theories, 
add the contribution of the adjoint chiral superfield, and 
take into account the condition on the global symmetry and R-symmetry charges imposed by the superpotential $W=Q\Phi Q$.
This was actually done for the $U(k)$ theory in \cite{Benini:2013xpa}, so we will follow their notation convention. 
We denote the holonomies of the global symmetry $Sp(n)\times U(1)_\Phi$ by $\xi_\alpha$, and $\lambda$, and that of the left-moving 
$U(1)$ R-symmetry by $z$.
As in \cite{Benini:2013xpa} we assign R-charge 0 to all fields but include the holonomy $\lambda$.
The superpotential will impose the constraint $\lambda = z$.

It will also be useful to label the flat connections $(U_1,U_2)^\pm$ by their exponents $u^\pm=(u_1^\pm,\cdots,u_{k}^\pm)$. 
In the $2\times 2$ blocks $e^{iu_{1i}\sigma_2},e^{iu_{2i}\sigma_2}$ with
continuous elements, the two associated $u$ parameters are given by
the two eigenvalues $\pm(u_{1i}+\tau u_{2i})$. In the blocks with discrete
numbers, we assign $u_i=0$ for an eigenvalue pair $(1,1)$,
$u_i=\frac{1}{2}$ for an eigenvalue pair $(-1,1)$, $u_i=\frac{\tau}{2}$ for $(1,-1)$, and $u_i=\frac{1+\tau}{2}$ for $(-1,-1)$. 
We therefore have
\be
\label{O(2p)HolonomiesAll}
u^+_{ab} = \left\{
\begin{array}{l}
(\pm u_1, \ldots, \pm u_p)   \\ 
(\pm u_1, \ldots, \pm u_{p-1},0,\frac{\tau}{2}) \\ 
(\pm u_1, \ldots, \pm u_{p-1},0,\frac{1}{2}) \\ 
(\pm u_1, \ldots, \pm u_{p-1},0,\frac{1+\tau}{2}) 
\end{array}
\right.
u^-_{ab} = \left\{
\begin{array}{l}
(\pm u_1, \ldots, \pm u_{p-2}, 0,\frac{1}{2}, \frac{1+\tau}{2}, \frac{\tau}{2}) \\ 
(\pm u_1, \ldots, \pm u_{p-1},\frac{1}{2}, \frac{1+\tau}{2})\\ 
(\pm u_1, \ldots, \pm u_{p-1},\frac{\tau}{2}, \frac{1+\tau}{2})\\ 
(\pm u_1, \ldots, \pm u_{p-1}, \frac{1}{2},\frac{\tau}{2})
\end{array}
\right.
\ee
for $O(2p)$, and
\be
\label{O(2p+1)HolonomiesAll}
u^+_{ab} = \left\{
\begin{array}{l}
(\pm u_1, \ldots, \pm u_p,0) \\
(\pm u_1, \ldots, \pm u_p,\frac{\tau}{2})\\
(\pm u_1, \ldots, \pm u_{p},\frac{1}{2})\\
(\pm u_1, \ldots, \pm u_{p},\frac{1+\tau}{2})
\end{array}
\right.
u^-_{ab} = \left\{
\begin{array}{l}
(\pm u_1, \ldots, \pm u_{p-1}, \frac{1}{2},\frac{1+\tau}{2}, \frac{\tau}{2})\\
 (\pm u_1, \ldots, \pm u_{p-1},\frac{1}{2}, \frac{1+\tau}{2},0)\\
(\pm u_1, \ldots, \pm u_{p-1},\frac{\tau}{2}, \frac{1+\tau}{2},0)\\
(\pm u_1, \ldots, \pm u_{p-1},\frac12,\frac{\tau}{2}, 0)
\end{array}
\right.
\ee
for $O(2p+1)$.

\subsection{$O(1)+n$ }
\label{$O(1)+n$}

In this case there is only the contribution of the matter fields:
\be
Z_{ab}^{O(1),n} =\frac{1}{2}\prod_{\alpha=1}^{n}\frac{\theta_1(-z\pm\xi_\alpha + \frac{a+b\tau}{2})}{\theta_1(\mbox{} \pm\xi_\alpha+\frac{a+b\tau}{2})} \,,
\ee
where the Jacobi's theta function definitions and the elliptic genus formulas are given in the appendix and we have also introduced the shorthand $\theta_1\big(A\pm B \big)\equiv \theta_1\big(A+B \big)\theta_1\big(A-B \big)$ 

\subsection{$O(2)+n$ }

The continuous sector is given by
\be\nonumber
Z_{00}^{O(2),n} &=& -\frac{1}{2} \frac{i\eta(q)^3}{\theta_1(-z)}\frac{\theta_1(\lambda-z)}{\theta_1(\lambda)}\oint_{u_*} du 
 \prod_{\alpha=1}^n
\frac{\theta_1(u- z \pm\xi_\alpha)}{\theta_1(u \pm\xi_\alpha)}
\frac{\theta_1(- u- z \pm\xi_\alpha)}{\theta_1(-u \pm\xi_\alpha)}\\\nonumber
&=& \sum_{\beta=1}^{n} \frac{\theta_1(2\xi_\beta \pm z)}{\theta_1(2\xi_\beta)^2}
\prod_{\alpha \neq \beta}^n
\frac{\theta_1(\xi_\beta\pm\xi_\alpha-z)}{\theta_1(\xi_\beta\pm\xi_\alpha)}
\frac{\theta_1(-\xi_\beta \pm\xi_\alpha-z)}{\theta_1(-\xi_\beta\pm\xi_\alpha)}
\ee
where we have picked up $2n$ residues at $u_*= \pm \xi_\beta$ and imposed the constraint $\lambda = z$.
In particular the $\theta_1(0)$ factor in the denominator cancels with the $\theta_1(\lambda - z)$ factor in the numerator.
The six discrete sectors are
\be\nonumber
Z_{ab\pm}^{O(2),n}&=&\frac{1}{4} \frac{\theta_1(u_{1ab}^\pm +u_{2ab}^\pm)}{\theta_1( -z + u_{1ab}^\pm+u_{2ab}^\pm )}
\frac{\theta_1(\lambda-z+u_{1ab}^\pm+u_{2ab}^\pm)}{\theta_1(\lambda+u_{1ab}^\pm+u_{2ab}^\pm)} 
\prod_{\alpha=1}^n\prod_{i=1}^2
\frac{\theta_1(u_{iab}^\pm- z \pm\xi_\alpha)}{\theta_1(u_{iab}^\pm \pm\xi_\alpha)}\nonumber\\
&=& \frac{1}{4} \frac{\theta_1(u_{1ab}^\pm +u_{2ab}^\pm)^2}{\theta_1( \pm z + u_{1ab}^\pm+u_{2ab}^\pm )}
\prod_{\alpha=1}^n\prod_{i=1}^2
\frac{\theta_1(u_{iab}^\pm- z \pm\xi_\alpha)}{\theta_1(u_{iab}^\pm \pm\xi_\alpha)}
\ee
with $(a,b)=(0,1), (1,0)$ and $(1,1)$, where $u_{iab}^\pm$ are given in (\ref{O(2p)HolonomiesAll}).

\subsection{$O(3)+n$}

There are four continuous sectors given by
\be
&&Z_{ab+}^{O(3),n} = -\frac{1}{4} \frac{i\eta(q)^3}{\theta_1(-z)} \oint_{u_*} du\Bigg( \frac{\theta_1(u+\frac{a+b\tau}{2})}{\theta_1(u-z+\frac{a+b\tau}{2})}\frac{\theta_1(-u+\frac{a+b\tau}{2})}{\theta_1(-u-z+\frac{a+b\tau}{2})}\\\nonumber
&&\times
\frac{\theta_1(u+\lambda - z+\frac{a+b\tau}{2})}{\theta_1(u+\lambda+\frac{a+b\tau}{2})}
\frac{\theta_1(-u+\lambda - z+\frac{a+b\tau}{2})}{\theta_1(-u+\lambda+\frac{a+b\tau}{2})} 
 \frac{\theta_1(\lambda - z+\frac{a+b\tau}{2})}{\theta_1(\lambda+\frac{a+b\tau}{2})}
\\\nonumber
&&\times 
 \prod_{i=\alpha}^{n}
\frac{\theta_1(u- z \pm\xi_\alpha)}{\theta_1(u \pm\xi_\alpha)}
\frac{\theta_1(-u- z\pm\xi_\alpha)}{\theta_1(-u \pm\xi_\alpha)}
\frac{\theta_1(-z\pm\xi_\alpha+\frac{a+b\tau}{2})}{\theta_1(\pm\xi_\alpha+\frac{a+b\tau}{2})}\Bigg) \,.
\ee
The JK poles are at $u_*=\big\{z-\frac{a+b\tau}{2},-\lambda-\frac{a+b\tau}{2}, \pm\xi_\beta\big\}$. 
Evaluating the residues is straightforward but cumbersome, so we will not show it here. 
The four discrete sectors are given by
\be\nonumber
Z_{ab-}^{O(3),n} &=& \frac{1}{8}  \prod_{i\neq j}^{3}\frac{\theta_1\big(u^-_{iab}+u^-_{jab}\big)}{\theta_1\big(-z +u^-_{iab}+u^-_{jab}\big)}
\frac{\theta_1(\lambda-z+u^-_{iab}+u^-_{jab})}{\theta_1(\lambda+u^-_{iab}+u^-_{jab})}
 \prod_{\alpha=1}^{n} \prod_{i=1}^{3}
\frac{\theta_1(-z +u^-_{iab}\pm\xi_\alpha)}{\theta_1(u^-_{iab}\pm\xi_\alpha)}\nonumber\\
&=& \frac{1}{8}  \prod_{i\neq j}^{3}\frac{\theta_1\big(u^-_{iab}+u^-_{jab}\big)^2}{\theta_1\big(\pm z +u^-_{iab}+u^-_{jab}\big)}
 \prod_{\alpha=1}^{n} \prod_{i=1}^{3}
\frac{\theta_1(-z +u^-_{iab}\pm\xi_\alpha)}{\theta_1(u^-_{iab}\pm\xi_\alpha)}
\ee
where $u^i_{ab-}$ are given in (\ref{O(2p+1)HolonomiesAll}).

\subsection{$\mathcal{N}=(4,4)$ duality}

Before we compare the partitions functions let us comment on their orbifold structure. 
Being a special case of the $(2,2)$-orthogonal theories, the $(4,4)-O(k)_\pm$ theories can also be obtained by gauging a $\mathbb{Z}_2$ global symmetry in the $SO(k)$ theory.
This symmetry is either charge conjugation, or charge conjugation combined with $(-1)^{F_s}$. The former is the so-called {\em standard orbifold}, and the latter, {\em non-standard} orbifold. In terms of the torus partition function this means 
\be
\label{standard Orbifold }
Z_{T^2}/\mathbb{Z}_2 &=& \frac12(Z_{00}+Z_{10}+Z_{01}+Z_{11}),\\\label{non standard Orbifold }
Z_{T^2}/\mathbb{Z}_2(-1)^{F_s} &=& \frac12(-Z_{00}+Z_{10}+Z_{01}+Z_{11}).
\ee
In \cite{Hori:2011pd}, the $O(k)_\pm$ theories where then defined  according to the parity of $N+k$ such that the three dualities always hold in the same form. For the $(4,4)$ theories since matter comes in hypermultiplets, we define (for any $k$) the $O(k)_+$ theory as the {\em standard orbifold} of $SO(k)$ and the $O(k)_-$ theory as the {\em non-standard orbifold} of $SO(k)$ (up to an overall sign).

The elliptic genus is then given by 
\be
\label{EGSOn}
Z^{SO(k),n}_{T^2}&=&2(Z_{00+}^{O(k),n} + y^{-2n}e^{i\theta} Z_{00-}^{O(k),n}), \\[5pt]
\label{EGOpn}
Z^{O(k)_+,n}_{T^2}&=&\sum_{a,b}(y^{-nb}Z_{ab+}^{O(k),n} + y^{n(b-2)}e^{i\theta} Z_{ab-}^{O(k),n})\\[5pt]
\label{EGOmn}
Z^{O(k)_-,n}_{T^2}&=&\sum_{a,b} (-1)^{ab+a+b+k} (y^{-nb}Z_{ab+}^{O(k),n} + y^{n(b-2)}e^{i\theta}Z_{ab-}^{O(k),n})
\ee
where the factors of $y = e^{2\pi i z}$ are required for modular invariance.
Setting $\theta = \pi$ for the regular theories, we have  analytically  showen (see appendix B) that the following equalities hold 
\be
Z^{SO(2),1}_{T^2}   &=& {Z}^{O(1)_+,1}_{T^2}, \\[5pt]
Z^{O(2)_+,1}_{T^2}  &=& {Z}^{SO(1),1}_{T^2}, \\[5pt]
Z^{O(2)_-,1} _{T^2} &=& {Z}^{O(1)_-,1}_{T^2}. 
\ee
For the higher rank dualities we have checked numerically (up to order $q^5$, where $q=e^{2\pi i \tau}$) that the following partition functions agree
\be
Z^{SO(3),2}_{T^2}   &=& {Z}^{O(2)_+,2}_{T^2}, \\[5pt]
Z^{O(3)_+,2}_{T^2}  &=& {Z}^{SO(2),2}_{T^2}, \\[5pt]
Z^{O(3)_-,2} _{T^2} &=& {Z}^{O(2)_-,2}_{T^2} \,,
\ee
confirming the proposed dualities in these cases.

In  the limit $z\rightarrow 0$ we obtain the Witten index, which is given by
\be
I_W^{SO(1),n} = I_W^{O(1)_-,n} = 1 \,, I_W^{O(1)_+,n} = 2 \,,
\ee
\be
I_W^{O(2)_+,n} = I_W^{O(2)_-,n} = n \,, I_W^{SO(2),n} = 2n \,,
\ee
and
\be
I_W^{SO(3),n} = I_W^{O(3)_-,n} = n \,, I_W^{O(3)_+,n} = 2n \,.
\ee
\subsection{Twisting}

\footnote{The following discussion is parallel for the (2,2) orthogonal theories defined in \cite{Hori:2011pd}.}  As mentioned in the previous section the  
The $O(k)_\pm$ theories are obtained by gauging the $SO(k)$ theory by a $\mathbb{Z}_2$ subgroup of its global symmetry that acts as charge conjugation. Their elliptic genera are therefore given by twisting the $SO(k)$ partition function and summing over the twists. The twisted $SO(k)$ elliptic genus is 
\be
Z^{SO(k),n}_{T^2(\alpha,\beta)}&=&2(Z_{\alpha\beta +}^{O(k),n} + y^{-2n}e^{i\theta} Z_{\alpha\beta-}^{O(k),n})
\ee
and the sums are given by (\ref{EGOpn}-\ref{EGOmn}).

The resulting orbifold theories $O(k)_\pm$ posses in turn a quantum  $\widehat{\mathbb{Z}}_2$ - global symmetry which acts non-trivially on the twisted(untwisted) sectors for the {\em standard}({\em non-standard}) orbifolds. Their twisted partition function is therefore given by
\be
Z^{O(k)_+,n}_{T^2(\alpha,\beta)}&=&\sum_{a,b}(y^{-nb}Z_{ab+}^{O(k),n} + y^{n(b-2)}e^{i\theta} Z_{ab-}^{O(k),n})(-1)^{\alpha b+\beta a},\\[5pt]\nonumber
Z^{O(k)_-,n}_{T^2(\alpha,\beta)}&=&(-1)^{k\alpha\beta}\sum_{a,b} (-1)^{ab+a+b+k} (y^{-nb}Z_{ab+}^{O(k),n} + y^{n(b-2)}e^{i\theta}Z_{ab-}^{O(k),n})(-1)^{\alpha (b+1)+\beta (a+1)}\\
\ee
where $(-1)^{k\alpha\beta}$ is a background term.

Setting  $ \theta=\pi $ we can now write a twisted generalization of the $(4,4)$-dualities 
\be
Z^{SO(k),n}_{T^2(\alpha,\beta)}  &=& {Z}^{O({2n-k+1})_+,n}_{T^2(\alpha,\beta)}, \\[5pt]
Z^{O(k)_+,n}_{T^2(\alpha,\beta)} &=& {Z}^{SO({2n-k+1}),n}_{T^2(\alpha,\beta)}, \\[5pt]
Z^{O(k)_-,n}_{T^2(\alpha,\beta)}  &=& {Z}^{O({2n-k+1})_-,n}_{T^2(\alpha,\beta)}. 
\ee
As a consistency check let us gauge the $\mathbb{Z}_2$ symmetry by summing over the twists. For example, for $k=n=1$ we find
\be
1/2\sum_{\alpha\beta}Z^{SO(1),1}_{T^2(\alpha,\beta)}&=&Z^{O(1)_+,1}_{(0,0)},\\[5pt]
 1/2\sum_{\alpha\beta}(-1)^{\alpha\beta+\alpha+\beta+1}Z^{SO(1),1}_{T^2(\alpha,\beta)}&=&Z^{O(1)_-,1}_{(0,0)},\\[5pt]
 1/2\sum_{\alpha\beta}Z^{O(1)_+,1}_{T^2(\alpha,\beta)}&=&Z^{SO(1),1}_{(0,0)},\\[5pt]
 1/2\sum_{\alpha\beta}(-1)^{\alpha\beta+\alpha+\beta}Z^{O(1)_+,1}_{T^2(\alpha,\beta)}&=&Z^{O(1)_-,1}_{(0,0)},\\[5pt]
 1/2\sum_{\alpha\beta}(-1)^{\alpha\beta+\alpha+\beta}Z^{O(1)_-,1}_{T^2(\alpha,\beta)}&=&Z^{O(1)_+,1}_{(0,0)},\\[5pt]
 1/2\sum_{\alpha\beta}(-1)^{\alpha\beta+\alpha+\beta+1}Z^{O(1)_-,1}_{T^2(\alpha,\beta)}&=&Z^{SO(1),1}_{(0,0)},
\ee
and for $k=2,n=1$ 
\be
1/2\sum_{\alpha\beta}Z^{SO(2),1}_{T^2(\alpha,\beta)}&=&Z^{O(2)_+,1}_{(0,0)},\\[5pt]
 1/2\sum_{\alpha\beta}(-1)^{\alpha\beta+\alpha+\beta}Z^{SO(2),1}_{T^2(\alpha,\beta)}&=&Z^{O(2)_-,1}_{(0,0)},\\[5pt]
 1/2\sum_{\alpha\beta}Z^{O(2)_+,1}_{T^2(\alpha,\beta)}&=&Z^{SO(2),1}_{(0,0)},\\[5pt]
 1/2\sum_{\alpha\beta}(-1)^{\alpha\beta+\alpha+\beta+1}Z^{O(2)_+,1}_{T^2(\alpha,\beta)}&=&Z^{O(2)_-,1}_{(0,0)},\\[5pt]
 1/2\sum_{\alpha\beta}(-1)^{\alpha\beta+\alpha+\beta+1}Z^{O(2)_-,1}_{T^2(\alpha,\beta)}&=&Z^{O(2)_+,1}_{(0,0)},\\[5pt]
 1/2\sum_{\alpha\beta}Z^{O(2)_-,1}_{T^2(\alpha,\beta)}&=&Z^{SO(2),1}_{(0,0)}.
\ee
Denoting the gauging operations $1/2\sum_{\alpha\beta}$ and  $1/2\sum_{\alpha\beta}(-1)^{\alpha\beta+\alpha+\beta+1}$ by $\mathbb{Z}_2$ and $\mathbb{Z}_2(-1)^{F_s}\equiv \mathbb{Z}_2'$ respectively and an overall sign by additional "-", we can summarize these relations with the following diagram
\be
\xymatrix{
SO(1) \ar@/_3pc/@{<->}[dd]_{\mathbb{Z}_2}\ar@{<->}[d]^{\mathbb{Z}_2'} \ar@{<=>}[r] &{O}(2)_+\ar@{<->}[d]_{\mathbb{Z}_2'}\ar@/^3pc/@{<->}[dd]^{\mathbb{Z}_2}\\
O(1)_- \ar@{<=>}[r]\ar@<0.4ex>[d]^{\mathbb{Z}_2} &{O}(2)_-\ar@<0.4ex>[d]^{\mathbb{Z}_2}\\
O(1)_+\ar@<0.4ex>[u]^{-\mathbb{Z}_2'} \ar@{<=>}[r] & {SO}(2)\ar@<0.4ex>[u]^{-\mathbb{Z}_2'}}
\ee
This is similar to Hori's $(2,2)$-orthogonal dualities  \cite{Hori:2011pd}, where starting from any duality the other two follow by the two gauging procedures. 

\section{Conclusions}

Using the elliptic genus machinery we investigated the brane based  duality conjuncture of $2d$ (4,4)-orthogonal gauge theories. Successfully testing the $O(1)\leftrightarrow O(2)$ and $O(2)\leftrightarrow O(3)$ cases. Our analysis reviles a similar structure to the (2,2)-dualities found in \cite{Hori:2011pd} where there are three theories and three pairs of dualities, all connected by gauging a global $Z_2$ symmetry. 

On the way we have classified the $O(k)$ flat connections on the torus. This has important application for any elliptic genus computation of orthogonal theories when the $\theta$ parameter of the theory is non-trivial.

These theories deserve further investigation, in particular it would be interesting to understand the mechanism that lifts the part of coulomb branch associated (in the (2,2) language) with the additional adjoint scalar.

\section{Acknowledgements}
We thank Hyungchul Kim, Jaemo Park, Shlomo Razamat, and Gabi Zafrir for useful discussions.
This work is supported in part by the Israel Science Foundation under grant no. 1390/17.

\appendix

\section{Rank one elliptic genus formula}
The elliptic genus of a gauge theory with gauge group $G$ of rank one is given by \cite{Benini:2013nda}
\be
Z_{T^2}= - \frac{1}{|W|} \sum_{u_j \in  M^+_{\text{sing}}} \oint_{u=u_j} \frac{i\eta(q)^3}{\theta_1(-z)} \prod_{\alpha \in G} \frac{\theta_1( \alpha u)}{\theta_1( \alpha u-z)}  \prod_{\Phi_i} \prod_{\rho\in \mathcal{R}_i} \frac{\theta_1 \big( (\frac{R_i}{2}-1) z+ \rho u +\xi_i\big)}{\theta_1 \big(\frac{R_i}{2} z+ \rho u+\xi_i \big)}
\ee
where $|W|$ is the order of the Weyl group, $\eta(q)$ is 
the Dedekind eta function
\be
\eta(q) = q^{1/24}\prod_{n=1}^\infty(1-q^n)
\ee
and $\theta_1$ is the Jacobi theta function
\be
\theta_1(\tau|z)= -i q^{\frac18} y^{\frac12} \prod_{k=1}^\infty (1-q^k)(1-yq^k)(1-y^{-1}q^{k-1}) 
\ee
with $q\equiv e^{2\pi i \tau}$ and $y\equiv e^{2\pi iz}$.
The first two terms of the integrand  are the contribution of gauge multiplet and the double product comes from the matter fields.
The integral is evaluated using the Jeffrey-Kirwan pole prescription
which for rank one gauge group amounts to taking the poles that are inside the unit circle consistently e.g if $u=z-\frac{a+b\tau}{2}\in  M^+_{\text{sing}}$ then $u=-z+\frac{a+b\tau}{2}$ does not.

\section{Playing with Jacobi theta functions}

In this appendix we derive the first three dualities  $Z^{S/O(2)_\pm,1}_{T^2}  = {Z}^{S/O(1)_\pm,1}_{T^2}$ using the known identities for the Jacobi theta functions. 

The Jacobi theta functions are defined as follows 
\be
&&\theta_1(\tau|z)= -i q^{\frac18} y^{\frac12} \prod_{k=1}^\infty (1-q^k)(1-yq^k)(1-y^{-1}q^{k-1}) ,\\
&&\theta_2(\tau|z)=  q^{\frac18} y^{\frac12} \prod_{k=1}^\infty (1-q^k)(1+yq^k)(1+y^{-1}q^{k-1}), \\
&&\theta_3(\tau|z)=  \prod_{k=1}^\infty (1-q^k)(1+yq^k)(1+y^{-1}q^{k-1}), \\
&&\theta_4(\tau|z)=  \prod_{k=1}^\infty (1-q^k)(1-yq^k)(1-y^{-1}q^{k-1}), 
\ee
they are known to satisfy many relations, we list here the ones that we need (we will use the shorthand $\theta_i(\tau|z)\equiv \theta_i(z), \theta_i(\tau|0)\equiv \theta_i$).\\

Shift symmetries:
\be
&&\theta_1(-z)=-\theta_1(z),\\[5pt]
&& \theta_1(z+a+b \tau)=(-1)^{a+b} e^{-2\pi ibz-\pi ib^2\tau}\theta_1(z).
\ee
\be
\label{relations1}
\begin{array}{ll}
\theta_1(z+\frac12)=\theta_2(z), &\quad \theta_4(z+\frac12)=\theta_3(z),\\[5pt]
\theta_1(z+\frac{1+\tau}{2})=q^{-\frac18}y^{-\frac12}\theta_3(z), &\quad \theta_4(z+\frac{1+\tau}{2})=q^{-\frac18}y^{-\frac12}\theta_2(z),\\[5pt]\label{relations2}
\theta_1(z+\frac{\tau}{2})=iq^{-\frac18}y^{-\frac12}\theta_4(z), &\quad \theta_4(z+\frac{\tau}{2})=iq^{-\frac18}y^{-\frac12}\theta_1(z).
\end{array}
\ee
Addition rules:
\be
\label{IdXpmY}
 &&\theta_1(x\pm z)=\frac{1}{ \theta_4^2}(\theta_1(x)^2\theta_4(z)^2-\theta_4(x)^2\theta_1(z)^2),\\[5pt]
 && \theta_2(x\pm z)=\frac{1}{ \theta_4^2}(\theta_2(x)^2\theta_4(z)^2-\theta_3(x)^2\theta_1(z)^2),\\[5pt]
 && \theta_3(x\pm z)=\frac{1}{ \theta_4^2}(\theta_3(x)^2\theta_4(z)^2-\theta_2(x)^2\theta_1(z)^2),\\[5pt]
 && \theta_4(x\pm z)=\frac{1}{ \theta_4^2}(\theta_4(x)^2\theta_4(z)^2-\theta_1(x)^2\theta_1(z)^2).
 \ee
Duplication formulas:
 \be
 \label{Id2X1}
&&  \theta_1(2x)=\frac{2\theta_1(x)\theta_2(x)\theta_3(x)\theta_4(x)}{  \theta_2 \theta_3\theta_4},\\[5pt] \label{Id2X4}
  &&  \theta_4(2x)=\frac{1}{\theta_4^3}(\theta_4(x)^4-\theta_1(x)^4) =\frac{1}{\theta_4^3}(\theta_3(x)^4-\theta_2(x)^4).
  \ee
Square identities:
  \be
  \label{1Idx^2}
&&\theta_1(x)^2\theta_4^2=\theta_3(x)^2\theta_2^2-\theta_2(x)^2\theta_3^2,\\[5pt]
&&\theta_2(x)^2\theta_4^2=\theta_4(x)^2\theta_2^2-\theta_1(x)^2\theta_3^2,\\[5pt]
&&\theta_3(x)^2\theta_4^2=\theta_4(x)^2\theta_3^2-\theta_1(x)^2\theta_2^2,\\[5pt]\label{2Idx^2}
&&\theta_4(x)^2\theta_4^2=\theta_3(x)^2\theta_3^2-\theta_2(x)^2\theta_2^2.
\ee

We can now derive the proposed dualities. Starting with  $Z^{SO(2),1}_{T^2}={Z}^{O(1)_+,1}_{T^2}$ we need to show that
\be
\frac14 \sum_{a,b=0}^1 \frac{\theta_1(\xi+\frac{a+b\tau}{2}\pm z)}{\theta_1(\xi+\frac{a+b\tau}{2})^2}-\frac{\theta
_1(2\xi\pm z)}{\theta_1(2\xi)^2}=0.
\ee
Using identity (\ref{IdXpmY}) we get
\be
 \frac14 \sum_{a,b=0}^1\Big(\frac{\theta_4(z)^2}{\theta_4^2} -\frac{\theta_1(z)^2\theta_4(\xi+\frac{a+b\tau}{2})^2}{\theta_4^2\theta_1(\xi+\frac{a+b\tau}{2})^2}\Big)-\Big(\frac{\theta_4(z)^2}{\theta_4^2} -\frac{\theta_1(z)^2\theta_4(2\xi)^2}{\theta_4^2\theta_1(2\xi)^2}\Big)=0,
\ee
so it is left to show that 
\be
 \frac14 \sum_{a,b=0}^1 \frac{\theta_4(\xi+\frac{a+b\tau}{2})^2}{\theta_1(\xi+\frac{a+b\tau}{2})^2}-\frac{\theta_4(2\xi)^2}{\theta_1(2\xi)^2}=0.
\ee
Using the relations (\ref{relations1}-\ref{relations2}) on the first term and (\ref{Id2X1}-\ref{Id2X4}) on the second we get
\be
\frac{1}{4}\Big(\frac{\theta_4(\xi)^2}{\theta_1(\xi)^2}+\frac{\theta_1(\xi)^2}{\theta_4(\xi)^2}+\frac{\theta_3(\xi)^2}{\theta_2(\xi)^2}+\frac{\theta_2(\xi)^2}{\theta_3(\xi)^2}\Big)-
\frac{\theta_2^2\theta_3^2}{4\theta_4^4}\frac{(\theta_4(\xi)^4-\theta_1(\xi)^4)^2}{\theta_1(\xi)^2\theta_2(\xi)^2\theta_3(\xi)^2\theta_4(\xi)^2}=0
\ee
or
\be\nonumber
(\theta_4(\xi)^4+\theta_1(\xi)^4)\theta_2(\xi)^2\theta_3(\xi)^2+(\theta_2(\xi)^4+\theta_3(\xi)^4)\theta_1(\xi)^2\theta_4(\xi)^2-
\frac{\theta_2^2\theta_3^2}{\theta_4^4}(\theta_4(\xi)^4-\theta_1(\xi)^4)^2=0,\\
\ee
finally using (\ref{1Idx^2}-\ref{2Idx^2}) we get an equality. \\

Next we show that $Z^{O(2)_+,1}_{T^2} = {Z}^{SO(1),1}_{T^2}$.
Starting with the LHS, we have the continuous sector 
\be
Z_{00}^{O(2),1}=\frac{\theta_1 (2\xi\pm z)}{\theta (2\xi)^2}
\ee
and the six discrete sectors which can be written in the following  form
\be
Z_{ab+}^{O(2),1}&=&\frac14\frac{\theta_1 (\xi\pm z) \theta_1(\frac{a+b \tau}{2})^2 \theta_1 (\xi\pm z+\frac{a+b \tau}{2})}{ \theta_1 (\xi )^2  \theta_1(z+\frac{a+b \tau}{2})^2\theta_1 (\xi+\frac{a+b \tau}{2})^2}\\[5pt]
Z_{ab-}^{O(2),1}&=&\frac14y^{(a+1)(1-b)}\frac{ \theta_1 (\frac{a+b \tau}{2})^2 \theta_1 (\xi\pm z+\frac{1-ab+\tau}{2}) \theta_1 (\xi\pm z+\frac{b+a(1-b)\tau}{2})}{\theta_1 (z+\frac{a+b\tau}{2})^2\theta_1 (\xi+\frac{1-ab+\tau)}{2})^2 \theta_1(\xi+\frac{b+a(1-b)\tau}{2})^2}
\ee
placing in (\ref{EGOpn}) and expanding the theta functions such that each contains only one fugacity (using  identities (\ref{IdXpmY}-\ref{Id2X4})) we end up with
\be\label{D1}
Z^{O(2)_+,1}_{T^2}&=&\frac{\theta_4(z)^2}{\theta_4^2} -
\frac{\theta_2^2\theta_3^2\theta_1(z)^2(\theta_4(\xi)^4-\theta_1(\xi)^4)^2}{4\theta_4^6\theta_1(\xi)^2\theta_2(\xi)^2\theta_3(\xi)^2\theta_4(\xi)^2}\\\nonumber
&+&\frac{\theta_2^2 \Big(\theta_1(\xi)^2 \theta_4(z)^2-\theta_4(\xi)^2 \theta_1(z)^2\Big) \Big(\theta_2(\xi)^2
   \theta_4(z)^2-\theta_3(x)^2 \theta_1(z)^2\Big)}{4 \theta_4^4 \theta_1(\xi)^2 \theta_2(\xi)^2 \theta_2(z)^2}\\\nonumber
    &-&\frac{\theta_2^2 \Big(\theta_4(\xi)^2 \theta_4(z)^2-\theta_1(\xi)^2 \theta_1(z)^2\Big) \Big(\theta_3(\xi)^2 \theta_4(z)^2-\theta_2(\xi)^2 \theta_1(z)^2\Big)}{4 \theta_4^4 \theta_3(\xi)^2 \theta_4(\xi)^2
   \theta_2(z)^2}\\\nonumber
   &+&\frac{\theta_3^2 \Big(\theta_1(\xi)^2 \theta_4(z)^2-\theta_4(\xi)^2 \theta_1(z)^2\Big)
   \Big(\theta_3(\xi)^2 \theta_4(z)^2-\theta_2(\xi)^2 \theta_1(z)^2\Big)}{4 \theta_4^4 \theta_1(\xi)^2
   \theta_3(\xi)^2 \theta_3(z)^2}\\\nonumber
    &-&\frac{\theta_3^2 \Big(\theta_4(\xi)^2 \theta_4(z)^2-\theta_1(\xi)^2 \theta_1(z)^2\Big) \Big(\theta_2(\xi)^2 \theta_4(z)^2-\theta_3(\xi)^2 \theta_1(z)^2\Big)}{4 \theta_4^4 \theta_2(\xi)^2 \theta_4(\xi)^2 \theta_3(z)^2}\\\nonumber
        &+&\frac{\Big(\theta_1(\xi)^2 \theta_4(z)^2-\theta_4(\xi)^2 \theta_1(z)^2\Big) \Big(\theta_4(\xi)^2 \theta_4(z)^2-\theta_1(\xi)^2 \theta_1(z)^2\Big)}{4 \theta_4^2 \theta_1(\xi)^2 \theta_4(\xi)^2 \theta_4(z)^2}\\\nonumber
&-&\frac{\Big(\theta_2(\xi)^2 \theta_4(z)^2-\theta_3(\xi)^2 \theta_1(z)^2\Big) \Big(\theta_3(\xi)^2 \theta_4(z)^2-\theta_2(\xi)^2 \theta_1(z)^2\Big)}{4 \theta_4^2 \theta_2(\xi)^2 \theta_3(\xi)^2 \theta_4(z)^2}
\ee
For the last step, comparing with the dual side 
\be\label{D2}
Z^{SO(1),1}_{T^2}=\frac{\theta_4(z)^2}{\theta_4^2}-\frac{\theta_4(x)^2\theta_1(z)^2}{\theta_4^2 \theta_1(x)^2}
\ee
we use \textit{Mathematica} to simplify the expression (\ref{D1}) minus (\ref{D2}) with the assumptions of identities (\ref{1Idx^2}-\ref{2Idx^2}), resulting with the expected zero.\\

The third duality $Z^{O(2)_-,1}_{T^2}={Z}^{O(1)_-,1}_{T^2}$ is also easy to prove following the same steps.

\end{document}